\documentclass{PoS}

\usepackage{amsmath}
\usepackage{epsfig}
\usepackage{graphicx}
\newlength{\bredde}
\def\slash#1{\settowidth{\bredde}{$#1$}\ifmmode\,\raisebox{.15ex}{/}
\hspace*{-\bredde} #1\else$\,\raisebox{.15ex}{/}\hspace*{-\bredde} #1$\fi}
\newcommand*{\chpt}{\raise0.4ex\hbox{$\chi$}PT}
\newcommand{\negcdot}{\negmedspace\cdot\negmedspace}

\newcommand{\Delstar}{\ensuremath{\Delta^{\raise0.18ex\hbox{${\scriptstyle *}$}}}}
\def\gtwid{{\,\raise.35ex\hbox{$>$\kern-.75em\lower1ex\hbox{$\sim$}}\,}}
\def\ltwid{{\,\raise.35ex\hbox{$<$\kern-.75em\lower1ex\hbox{$\sim$}}\,}}
\def\leftvec{{\raise1.5ex\hbox{$\leftarrow$}\kern-1.00em}}
\def\rightvec{{\raise1.5ex\hbox{$\rightarrow$}\kern-1.00em}}
\def\half{{\scriptstyle \raise.2ex\hbox{${1\over2}$}}}
\def\threehalves{{\scriptstyle \raise.15ex\hbox{${3\over2}$}}}
\def\third{{\scriptstyle \raise.15ex\hbox{${1\over3}$}}}
\def\third{{\scriptstyle \raise.15ex\hbox{${1\over3}$}}}
\def\twothirds{{\scriptstyle \raise.15ex\hbox{${2\over3}$}}}
\def\fourth{{\scriptstyle \raise.15ex\hbox{${1\over4}$}}}
\def\la{\langle}
\def\ra{\rangle}

\def\op{{\mathcal{O}}}

\def\exponential{{\mathrm{e}}}

\def\calP{{\mathcal{P}}}

\def\calV{{\mathcal{V}}}

\def\beq{\begin{equation}}
\def\eeq{\end{equation}}
\def\barr{\begin{array}}
\def\earr{\end{array}}

\newcommand{\cM}{\ensuremath{\mathcal{M}}}

\newcommand{\cO}{\ensuremath{\mathcal{O}}}

\newcommand{\qbar}{\ensuremath{\overline{q}}}

\newcommand{\Dbar}{\ensuremath{\overline{D}}}

\newcommand{\cbar}{\ensuremath{\overline{c}}}
\newcommand{\ubar}{\ensuremath{\overline{u}}}

\newcommand*{\bea}{\begin{eqnarray}}
\newcommand*{\eea}{\end{eqnarray}}
\newcommand*{\be}{\begin{equation}}
\newcommand*{\ee}{\end{equation}}

\title{Chiral expansion for lattice computations\\of $B^{+}\to D^{0}K^{+}(\pi^{+})$ and 
       $B^{+}\to \Dbar^{0} K^{+}(\pi^{+})$ amplitudes}

\ShortTitle{$B\rightarrow DK$}

\author{Christopher Aubin$^{a,\$ }$, 
     C.-J.~David~Lin$^{b,c,\dagger}$, Amarjit~Soni$^{d,\ddagger}$\\
$^{a}$ Department of Physics, Fordham University, Bronx, NY 10458, USA\\
$^{b}$ Institute of Physics, National Chiao-Tung University, Hsinchu  300, Taiwan\\
$^{c}$ Division of Physics, National Centre for Theoretical Sciences, Hsinchu 300, Taiwan\\
$^{d}$ Physics Department, Brookhaven National Laboratory, Upton, NY
11973, USA\\ \ \\
       $^{\$}$E-mail: \email{caubin@fordham.edu}\\
       $^{\dagger}$E-mail: \email{dlin@mail.nctu.edu.tw}\\
       $^{\ddagger}$E-mail: \email{soni@quark.phy.bnl.gov}}

\abstract{In this work, we suggest that hard-pion chiral perturbation
  theory may be applicable to
the real parts of nonleptonic $B^{+}\to D^{0}P^{+}$ and $B^{+}\to \overline{D}^{0}P^{+}$ ($P=K,\pi$) 
decay amplitudes.  These amplitudes play an important role in the extraction of the 
angle $\gamma$ in the $b{-}d$ unitarity triangle of the CKM matrix, and their real parts can
be computed using lattice QCD. We construct the leading-order
operator in the chiral expansion for these nonleptonic decays, and discuss the generic features
of the next-to-leading-order terms.}

\FullConference{ The XXIX International Symposium on Lattice Field Theory - Lattice 2011\\
July 10-16, 2011\\
Squaw Valley, Lake Tahoe, California}

\begin{document}

\section{Introduction}
\label{sec:intro}
Nonleptonic $B$ decays have played an important role in CKM physics.
In particular, the angle $\gamma$ in the $b{-}d$ unitarity triangle can in principle be determined
with high precision from the study of charged $B$ meson decays processes,
\bea
 B^{+} \to D^{0} P^{+} \to f P^{+}\mbox{ }{\mathrm{and}}\mbox{ }\mbox{ } B^{+} \to \Dbar^{0} P^{+} \to f P^{+},
\eea
with the same final state $f P^{+}$~\cite{Gronau:1990ra,Gronau:1991dp,Atwood:1996ci,hep-ph/0008090}.  
In all the methods for extracting $\gamma$ from the above decay channels, the knowledge of the 
branching ratios, $Br[B^{+} \to D^{0} P^{+}]$ and $Br[B^{+} \to \Dbar^{0} P^{+}]$, is a key
ingredient.  While $Br[B^{+} \to D^{0} P^{+}]$ has been experimentally measured with good accuracy, 
$Br[B{^+} \to \Dbar^0 P{^+}]$ is very difficult to obtain.
For this reason, despite the large statistics of the two $B$-factories 
[$\sim\cO(10^9)$ charged $B$ meson samples], $\gamma$ is presently determined to 
only $\sim\cO(25\%)$. This is compared with 
about 3\% for $\beta$, and about 5\% for $\alpha$. 
To further improve the precision on  $\gamma$, inputs from lattice QCD (LQCD) to these branching ratios
will be very helpful

\smallskip

The study of hadronic weak decays on the lattice is very challenging, because of the 
Maiani-Testa no-go theorem (MTNGT) \cite{Maiani:1990ca}.   This theorem states
that by using Euclidean four-point correlators to study nonleptonic two-body decays, it is
impossible to obtain information about the strong phases.  Therefore one can
only compute the real parts of nonleptonic decay amplitudes from
such correlators.  For the calculation of $K \to \pi \pi$ on the
lattice, one can avoid the MTNGT using the 
Lellouch-L\"{u}scher (LL) method \cite{Lellouch:2000pv}.   On the other hand, the
lattice computation of nonleptonic $B$ decays remains challenging,
because the LL method is only applicable to processes involving elastic
final-state scatterings. Nevertheless, lattice results for the real part of these amplitudes could provide valuable
information of $Br[B^{+} \to D^{0} P^{+}]$ and $Br[B^{+} \to \Dbar^{0} P^{+}]$, and help in the extraction of $\gamma$.

\smallskip

In the near future, lattice computations for such four-point functions
are unlikely to be preformed exactly at physical pion
mass.  Thus, it is important to rely on the chiral expansion to perform extrapolations to the
physical point.
In Ref.~\cite{Aubin:2011kv},  we examine the possibility of the chiral expansion of the real parts
of $B^{+}\to D^{0} P^{+} $ and $B^{+}\to \Dbar^{0} P^{+}$ amplitudes,
in the framework of heavy-meson chiral perturbation theory
(HM\chpt).
\cite{Wise:1992hn,Burdman:1992gh,Yan:1992gz,Cho:1992gg,Cho:1992cf}.
The straightforward application of the chiral expansion for $B\to DP$ amplitudes near the 
physical kinematics is questionable, since the final-state hadrons carry momenta
$\sim 2$ GeV.  From the structure of subleading terms in $1/M_{D}$
($M_{D}$ is the $D$ meson mass) in the effective theory
\cite{Boyd:1994pa,Stewart:1998ke,Aubin:2005aq}, it is clear that the large momentum
carried by the $D$ meson will only lead to (significant) dependence on
$M_{D}$ in the low-energy constants (LEC's).   What remains to be considered is
the convergence of the chiral expansion for hard final-state hadrons
from lattice studies.

\smallskip

In Ref.~\cite{Aubin:2011kv}, we address this issue in nonleptonic $B$ decays,
in a framework that is similar to the hard-pion chiral perturbation
theory (HP$\chi$PT) which was first proposed and applied to the analysis of 
$K_{\ell3}$ decays in Ref.~\cite{Flynn:2008tg}.
This approach was also used to carried out investigations of
$K\to  2\pi$ amplitudes~\cite{Bijnens:2009yr}\footnote{As pointed out in 
Ref.~\cite{Bijnens:2009yr}, hard-pion $\chi$PT may not be used to
estimate the strong phases in 
$K\to \pi\pi$ processes.  We will comment on this issue for nonleptonic $B$ decays 
in Sec.~\ref{sec:hardPi}.}  ,
as well as semi-leptonic $B$-decays~\cite{Bijnens:2010ws}. 

\smallskip

The relevant current-current, $\Delta b=1$, operators ($\alpha,\beta$ are colour indices)
in our work are,
\bea
	Q^{b\to c,i}_1 &=&
	 (\qbar^i_\alpha \gamma^\mu(1-\gamma_5)b_\alpha)
	(\cbar_\beta\gamma_\mu(1-\gamma_5) u_\beta)\ ,\label{eq:btoc1}\\
	Q^{b\to c,i}_2 &=&
	 (\qbar^i_\alpha\gamma^\mu(1-\gamma_5) b_\beta)
	(\cbar_\beta \gamma_\mu(1-\gamma_5)u_\alpha)\ ,\label{eq:btoc2}\\
	Q^{b\to \cbar,i}_1 &=&
	(\qbar^i_\alpha \gamma^\mu(1-\gamma_5)b_\alpha)
	(\ubar_\beta \gamma_\mu(1-\gamma_5)c_\beta)\ ,
	\label{eq:btocbar1}\\
	Q^{b\to \cbar,i}_2 &=&
	(\qbar^i_\alpha\gamma^\mu(1-\gamma_5) b_\beta)
	(\ubar_\beta \gamma_\mu(1-\gamma_5)c_\alpha)\ .\label{eq:btocbar2}
\eea
We will focus on the nonleptonic decays with the underlying processes 
$b \to c\ubar d,\ b \to c\ubar s,\ b \to u\cbar d,$ 
and $b \to u\cbar s$. The first two will be mapped onto different operators in 
the chiral effective theory from the last two, as explained in the next section.

\section{The leading-order chiral expansion for $B \to D K (\pi)$ amplitudes}
\label{sec:ChPT_BtoDK}
In weak nonleptonic $B$ decays, the final-state hadrons carry large momenta.  This makes
it difficult to apply the chiral expansion to such processes.
On the other hand, it has been recently proposed that chiral
perturbation theory ($\chi$PT) can be valid for amplitudes 
containing hard final-state particles~\cite{Flynn:2008tg,Bijnens:2009yr,Bijnens:2010ws,Bijnens:2010jg}. 
One crucial point is to allow the low-energy constants (LECs) to depend on the hard momentum scales which 
result from either the kinematics or the mass of the external particles.  That is, the LECs in
the chiral expansion are no longer universal quantities.  Another important
ingredient in such chiral expansions is the separation of the hard and soft scales.  This separation of
scales is made possible because of the structure of derivative couplings in $\chi$PT, leading to the 
absorption of the hard, external, momenta into the LECs. We will discuss this procedure explicitly for 
$B\to DP$ amplitudes with an example diagram in Sec.~\ref{sec:hardPi}.

\smallskip

In this section, we construct the \chpt\ weak operators corresponding to those in 
Eqs.~(\ref{eq:btoc1})--(\ref{eq:btocbar2}).
%(\ref{eq:btoc2}), (\ref{eq:btocbar1}) and (\ref{eq:btocbar2}).  
Neglecting the colour indices which do not play a role in \chpt, these operators can be written as
\bea
     && Q^{b\to c,i}
	= \left ( \qbar^{i}_{L} \mbox{ }\Gamma_{1}\mbox{ } b \right ) \left ( \cbar \mbox{ }\Gamma_{2}\mbox{ } u_{L} \right ) , \nonumber\\
&&\nonumber\\
\label{eq:Q_chiral_form}
     && Q^{b\to \overline{c},i}
	= \left ( \qbar^{i}_{L} \mbox{ } \overline{\Gamma}_{1} \mbox{ } b \right ) \left ( \ubar_{L} \mbox{ } \overline{\Gamma}_{2}\mbox{ } c \right ) ,
\eea
where $q^{i} = d$ or $s$, and 
\beq
  q_{L} = \left ( \frac{1 - \gamma_{5}}{2} \right )\mbox{ } q ;
\mbox{ }\mbox{ }
 \Gamma_{1} = \Gamma_{2}  =  \overline{\Gamma}_{1} = \overline{\Gamma}_{2} = \gamma_{\mu} (1 - \gamma_{5}) .
\eeq
Under the ${\mathrm{SU}}(3)_{{\mathrm{L}}} \otimes {\mathrm{SU}}(3)_{{\mathrm{R}}}$
chiral symmetry group, $Q^{b\to c,i}$ is in the $({\bf 8_L,1_R})$ representation, while $Q^{b\to \overline{c},i}$ is in the 
$({\bf \bar{6}_L,1_R})$ representation.  To bosonise these operators, we promote $\Gamma_{1,2}$ and $\overline{\Gamma}_{1,2}$ to
be spurion fields which transform as
\beq
\label{eq:spurion_transformation}
 \Gamma_{1} \to L\mbox{ }\Gamma_{1}\mbox{ }S^{\dagger} , \mbox{ }\mbox{ }
    \Gamma_{2} \to S\mbox{ }\Gamma_{2}\mbox{ }L^{\dagger} ,
\mbox{ }\mbox{ }
 \bar{\Gamma}_{1} \to L\mbox{ }\overline{\Gamma}_{1}\mbox{ }S^{\dagger} , \mbox{ }\mbox{ }
     \bar{\Gamma}_{2} \to L\mbox{ }\overline{\Gamma}_{2}\mbox{ }S^{\dagger} ,
\eeq
under the heavy-quark spin/flavour and chiral rotations. This renders the 
operators in Eq.~(\ref{eq:Q_chiral_form}) invariant with respect to such transformations. 
Using this property, and taking into account all the possible insertions of 
Dirac structures~\cite{Detmold:2006gh},
we obtain the leading-order (LO) operators,
\bea
 \op_{\chi,i} = &&
  \left [ \beta_{1} + 
   \left ( \beta_{1} + \beta_{2}\right ) (v^{\prime}\cdot v)
  \right ] 
  \left [ \left ( \sigma_{1k}  \calP^{(c)\dagger}_{k}\right ) 
          \left ( \calP^{(b)}_{l} \sigma^{\dagger}_{li} \right )
  \right ]
\nonumber\\
&+&
  \left [ \left ( \beta_{1} - \beta_{2}\right ) v^{\prime\mu}
          - \beta_{1} v^{\mu}
  \right ] 
  \left [ \left ( \sigma_{1k}  \calP^{(c)\dagger}_{k}\right ) 
          \left ( \calV^{\ast(b)}_{\mu,l} \sigma^{\dagger}_{li} \right )
  \right ]\nonumber\\
  &+&
  \left [ \beta_{1} v^{\prime\mu}
          - \left ( \beta_{1} + \beta_{2}\right ) v^{\mu}
  \right ] 
  \left [ \left ( \sigma_{1k}  \calV^{\ast(c)\dagger}_{\mu,k}\right ) 
          \left ( \calP^{(b)}_{l} \sigma^{\dagger}_{li} \right )
  \right ]\nonumber\\
  &-& 4
  \left [ \left ( \beta_{1} - \beta_{2}\right )
         + \beta_{1} (v^{\prime}\cdot v)
  \right ] 
  \left [ \left ( \sigma_{1k}  \calV^{\ast(c)\dagger}_{\mu,k}\right ) 
          \left ( \calV^{\ast(b)\mu}_{l} \sigma^{\dagger}_{li} \right )
  \right ], \nonumber\\
&&\nonumber\\
 \overline{\op}_{\chi,i} = &&
  \left [ \overline{\beta}_{1} + \overline{\beta}_{2} (v^{\prime}\cdot v)
  \right ] 
  \left [ \left ( \calP^{(\bar c)\dagger}_{k} \sigma_{k1}^{\dagger}  \right ) 
          \left ( \calP^{(b)}_{l} \sigma^{\dagger}_{li} \right )
  \right ]\nonumber\\
  &-&
  \left [ \overline{\beta}_{2} v^{\prime\mu} - \left ( \overline{\beta}_{1} + \overline{\beta}_{5} \right ) v^{\mu}
       - \overline{\beta}_{3} (v^{\prime}\cdot v) v^{\mu}
  \right ] 
  \left [ \left (\calP^{(\bar c)\dagger}_{k}  \sigma_{k1}^{\dagger} \right ) 
          \left ( \calV^{\ast(b)}_{\mu,l} \sigma^{\dagger}_{li} \right )
  \right ]\nonumber\\
  &+&
  \left [ \overline{\beta}_{1} v^{\prime\mu} - \overline{\beta}_{2} v^{\mu}
  \right ] 
  \left [ \left (  \calV^{\ast(\bar c)\dagger}_{\mu,k} \sigma_{k1}^{\dagger} \right ) 
          \left ( \calP^{(b)}_{l} \sigma^{\dagger}_{li} \right )
  \right ]\nonumber\\
\label{eq:chiral_O_expanded}
  &+& 
  \left [ 4 \overline{\beta}_{2} - \overline{\beta}_{3}
     - 2 \left ( \overline{\beta}_{1} + \overline{\beta}_{4} +\overline{\beta}_{5} \right ) 
      (v^{\prime}\cdot v)
  \right ] 
  \left [ \left (  \calV^{\ast(\bar c)\dagger}_{\mu,k} \sigma_{k1}^{\dagger} \right ) 
          \left ( \calV^{\ast(b)\mu}_{l} \sigma^{\dagger}_{li} \right )
  \right ], 
\eea
where $\beta_{i}$ and
$\overline{\beta}_{i}$ are LECs, $v$ and $v^{\prime}$ are the velocities of $B$ and $D$ mesons,
$\sigma = \sqrt{\Sigma} = $exp$(i\Phi/f)$ with $\Phi$ and $\Sigma$ being the standard linear and nonlinear
Goldstone fields, $\calP^{(h)}$ and $\calV^{\ast(h)}$ are the anihilation fields for the pseudoscalar and vector 
mesons containing the heavy quark $h$.
At the LO,
% it is straightforward to demonstrate, using Eq.~(\ref{eq:chiral_O_expanded}),
%
\bea
 && \la D^{0} K^{-} | \op_{\chi,s} | B^{-}\ra = \la D^{0} \pi^{-} | \op_{\chi,d} | B^{-}\ra
      = \frac{i}{f} \la D^{-} | \op_{\chi,s} | B^{-}\ra ,\nonumber\\
&& \nonumber\\
\label{eq:LO_chiral_expansion}
 && \la \overline{D}^{0} K^{-} | \overline{\op}_{\chi,s} | B^{-}\ra = \la \overline{D}^{0} \pi^{-} | \overline{\op}_{\chi,d} | B^{-}\ra
      = \frac{i}{f} \la D^{-} | \overline{\op}_{\chi,s} | B^{-}\ra 
      \ .
\eea
This is not valid at the next-to-leading-order (NLO) chiral expansion.

%\begin{figure}[t]%[htbp]
%\begin{center}
%\includegraphics[width=5in]{TreeLevel1.eps}
%\caption{Tree-level diagrams contributing to (a) $B\to D$ and (b) $B\to DP$ at lowest order, with no insertions of the strong Lagrangian. The box is the weak operator, the solid line is a heavy-light pseudoscalar (either $B$ or $D$), and the dashed line is the light meson $P$.}
%\label{fig:BtoDtreeOrderOne}
%\end{center}
%\end{figure}

\section{Resonance contributions}
\label{sec:resonance}

In this section we investigate resonance contribution to $B\to D P$ 
correlators (in the time-momentum representation) and amplitudes\footnote{The conclusion
presented in this section is also valid for $B\to \overline{D} P$ decays.}. 
The vector heavy-light resonances are already incorporated in
HM$\chi$PT. 
The inclusion of heavier resonances in the effective
theory is beyond the scope of this work.  
Because of the complications in formulating HM\chpt\ in Euclidean
space, and here we are only sutdying the generic feature of the
resonance contribution,
we work in Minkowski space.

\smallskip
\begin{figure}[t]%[htbp]
\begin{center}
\includegraphics[width=4in,height=0.8in]{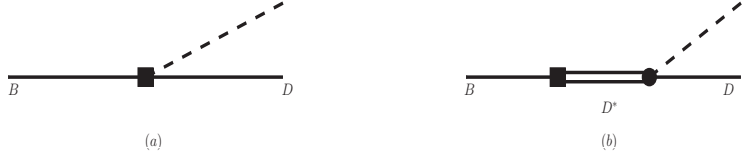}
\caption{Tree-level correlators contributing to $B\to DP$. The box is the weak operator, with (a) being the direct $B\to DP$ term and (b) being the term with an intermediate resonance (here a $D^*$).}
\label{fig:BtoDtreeOrderPpi}
\end{center}
\vspace{-0.7cm}
\end{figure}
First, we calculate the LO correlator in
Fig.~\ref{fig:BtoDtreeOrderPpi}(a).  We perform a Fourier transform
for the spatial directions for each of the external-source points,
while fixing the location of the weak operator (the square in the
diagram) to be at the origin.  As in Sec.~\ref{sec:ChPT_BtoDK},
we denote the velocity of $B$ and $B^{\ast}$ by $v$ and that of 
$D$ and $D^{\ast}$ by $v^{\prime}$.  For simplicity, we choose the
frame in which $v = \left ( 1, \vec{0}\right )$, and implement 
the time-ordering as $ t_{B} < 0 < t_{D} \le t_{P}$,
where $t_{B,D,P}$ are the temporal locations of the sources for $B$,
$D$ and $P$ mesons. 
Using the the weak operators 
in Eq.~(\ref{eq:chiral_O_expanded}), the result of this diagram is 
\beq
\label{eq:LO_correlator}
 C_{{\mathrm{LO}}}  =\frac{g_{BDP}}{f} \left ( \frac{1}{2} \right ) 
 \left ( \frac{\exponential^{-i \overline{\delta}_{D} t_{D} } }{2 v^{\prime}_{0}}\right )  
 \left ( 
   \frac{\exponential^{-i \omega_{P} t_{P}}}{2 \omega_{P}} \right ) ,
\eeq
where
$\overline{\delta}_{D} = \vec{v^{\prime}} \cdot \vec{p}_{D},
 \mbox{ }\mbox{ }{\mathrm{and}}\mbox{ }\mbox{ } 
 \omega_{P} = \sqrt{M^{2}_{P} + \vec{p}^{2}_{P}}$,
with $\vec{p}_{D}$ and $\vec{p}_{P}$ denoting the spatial momenta of
the $D$ and the $P$ mesons.
The coupling $g_{BDP}$ is one of the linear combinations of the LEC's $\beta_{i}$ in 
Eq.~(\ref{eq:chiral_O_expanded}). 

\smallskip

Next, we compute the diagram in Fig.~\ref{fig:BtoDtreeOrderPpi}(b),
which leads to the result
\beq
\label{eq:resonance_correlator}
 C_{{\mathrm{res}}} = \frac{g_{BD^{\ast}} g_{\pi} }{f^{2}}\left ( \frac{1}{2} \right ) 
 \left ( \frac{\exponential^{-i \overline{\delta}_{D} t_{D} }}{2 v^{\prime}_{0}}  \right )  
 \left ( 
   \frac{\exponential^{-i \omega_{P} t_{P}}}{2 \omega_{P}} \right )
 \left [
  \frac{\exponential^{i (\omega_{P} + \overline{\delta}_{D} - \overline{\Delta}_{DP} ) t_{D}} - 1}
   { 2  v^{\prime}_{0} (\omega_{P} + \overline{\delta}_{D} - \overline{\Delta}_{DP} )}
 \right ] ,
\eeq
where $
 \overline{\Delta}_{DP} = \vec{v}^{\prime} \cdot \left ( \vec{p}_{D} + \vec{p}_{\pi} \right )
     + \Delta_{D} /v^{\prime}_{0}$ ,
with $\Delta_{D}$ denoting the $D^{\ast}{-}D$ mass splitting resulting from the heavy-quark
spin symmetry breaking effects.  
The coupling $g_{\pi}$ is the LO $B^{\ast}{-}B{-}\pi$ axial coupling in
the HM$\chi$PT Lagrangian, and $g_{BD^{\ast}}$ is a linear combination of the LEC's $\beta_{i}$ in
Eq.~(\ref{eq:chiral_O_expanded}).  Notice that $g_{BD^{\ast}}$ is different from $g_{BDP}$ and thus the
resonance contribution results in general in an additional unknown parameter for $B\to D P$ amplitude 
at the tree level.
When the final-state momenta
are tuned such that
the resonance is on-shell,  the factor in the square brackets in 
Eq.~(\ref{eq:resonance_correlator}) 
becomes linear in $t_{D}$.  This can be interpreted as an energy shift of
the final state.   In any case, the $t_{D}$ dependence in these square
brackets will be cancelled when taking the ration between
$C_{{\mathrm{LO}}}+C_{{\mathrm{res}}}$ and the square root of the $D P
\to D P$ correlator to obtain the $B\to DK$ matrix element.

\section{Hard-pion chiral perturbation theory at one-loop}
\label{sec:hardPi}
To account for nonanalytic dependence on $m_{\pi}$ in $B\to D K$
amplitudes, it is necessary to go beyond the tree-level in the chiral
expansion.  In order to treat these processes in the physical regime,
we use the methods of
Refs.~\cite{Flynn:2008tg,Bijnens:2009yr,Bijnens:2010ws}: Hard-pion
\chpt\ (HP\chpt), in which one key point is the
separation of the hard external momenta with the soft scales (the
Goldstone masses) which appear in the loop.  This is acheivable
because of the structure of the derivative couplings.
Focusing on the SU(2) $\chi$PT, the chiral logarithms for $B\to DK$
amplitudes in HP$\chi$PT takes the generic form,
\be\label{eq:Moneloop}
	\cM = \cM^{\rm tree}\left[1 
	+ a\frac{m_\pi^2}{16\pi^2 f^2}
	\ln \left(\frac{m_\pi^2}{\Lambda^2}\right)
	+ L m_\pi^2
	\right] \ ,
\ee
where $\cM^{\rm tree}$ is its LO value. $a$ is a coefficient that
depends on the particular kinematics chosen for the diagram, and $L$
is a linear combination of low-energy constants as well as terms
arising from higher-order chiral-level weak operators. The coefficient
$a$ would be determined from evaluating the one-loop corrections to
the amplitude. Both $a$ and $L$ depend 
on all of the hard quantities (such as the momenta of the external $D$ meson and pion).

\smallskip

\begin{figure}[t]%[htbp]
\begin{center}
\includegraphics[width=2.5in,height=0.6in]{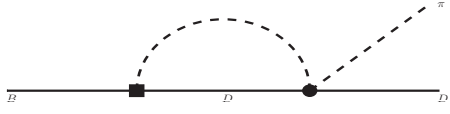}
\caption{One of the many one-loop diagrams that contribute to $B\to D\pi$, specifically one which shows the essential features that arise in HP\chpt.}
\label{fig:Sunset1_BtoDpi}
\end{center}
\vspace{-0.7cm}
\end{figure}
In order to understand the specific details, we examine the diagram in
Fig.~\ref{fig:Sunset1_BtoDpi}, through which we will demonstrate the
dependence on the hard scales in the coefficient of the chiral logarithm.
In this diagram, as in lattice calculations, momentum will be
conserved at the strong vertex, but need not be at
the weak operator.  This momentum insertion, $p_{\rm wk}$, is related
to the momenta of the external mesons via
$p_{B} + p_{\rm wk} = p_{D} + p_{\pi} $.
Denoting the residual momentum of the external $D$ by $k$, and the
internal pion momentum by $l$, the integral that needs to be computed
for this diagram is 
\beq
I  = \int
	\frac{d^4\ell}{(2\pi)^4}
	\frac{i}{\ell^2 - m_\pi^2 + i\epsilon}
	\frac{iv'\cdot(\ell-p_\pi)}{v'\cdot(\ell - k - 
	p_\pi)-\Delta+i\epsilon}\ ,
	\label{eq:BtoDpi1loop}
\eeq
where $v^{\prime}$ is the velocity of the $D$ mesons, and 
$\Delta = m_D - m_B$ is the $D$-$B$ meson mass splitting.
This integral can be evaluated using dimensional regularisation,
\be
I =
	\frac{1}{16\pi^2f^2}\left[
	\frac{v'\negcdot k+\Delta}{v'\negcdot (k+p_\pi)+\Delta+i\epsilon}I_2(m_\pi, v'\negcdot (k+p_\pi)+\Delta+i\epsilon)
	-m_\pi^2 \ln\left(\frac{m_\pi^2}{\Lambda^2}\right)\right]
	\ ,
\ee
where $I_{2}$ takes the form in the hard-pion limit $v'\negcdot k\gg m_\pi$, 
\be
	I_2(m_\pi, v'\negcdot (k+p_\pi)+\Delta) \approx
	-m_\pi^2\ln\left(\frac{m_\pi^2}{\Lambda^2}\right)\ ,
%	2(v'\negcdot k-\Delta)^2}\right)
\ee
so that the full coefficient $a$ [as defined in
Eq.~(\ref{eq:Moneloop}) ] in the integral $I$ is $-2$ when we
insert momentum into the weak vertex such that $p_\pi\approx 0$, and
is $-3/2$ when we choose $p_{\rm wk}$ such that $p_\pi\approx k$. 
Here we comment that the imaginary part in this diagram is
proportional to $\sqrt{[v^{\prime}\cdot (p_{\pi}+k)]^{2} -
  m^{2}_{\pi}}$, therefore grows with the final-state momenta, leading to the failure of the chiral expansion when $p_{D}$ and $p_{\pi}$ are large.  This can be understood by noting that the imaginary part arises from the contribution in which both mesons in the loop are on-shell, and therefore cannot be soft.

\section{Summary}
We studied the chiral expansion for lattice
computations of the real parts of the $B^{+}\to
D^{0} P^{+}$ and $B^{+}\to \overline{D}^{0}P^{+}$, which are important
inputs in the determination of the angle $\gamma$ in the $b{-}d$
untarity triangle in the CKM matrix.  The calculation of these real
parts are not obstacled by the MTNGT.   We derived the LO operators
for these matrix elements in HM$\chi$PT, argue that HP$\chi$PT is
applicable to these computations, and investigate some generic features
of the NLO contributions.

\section*{Acknowledgments}
This work has been supported by 
the US DOE under grant number DE-AC02-98CH10886 and Taiwanese NSC grant 99-2112-M-009-004-MY3.  
C.-J.D.L. thanks the hospitality of Fordham University and
Brookhaven National Lab during the progress of this work.

\end{document}